\documentstyle[12pt]{article}
\begin{document}
\begin{center}
{\large\bf EXOTIC $\rho^\pm\rho^0$ STATE PHOTOPRODUCTION} \\[2cm]
{\large N.N. Achasov and G.N. Shestakov}\\[0.6cm]
{\it Laboratory of Theoretical Physics,
S.L. Sobolev Institute for Mathematics,\\[1pt]
630090, Novosibirsk 90, Russia}\\[2.5cm]
Abstract\\[0.5cm] \end{center}

It is shown that the list of unusual mesons planned for a careful study in
photoproduction can be extended by the exotic states $X^\pm(1600)$ with $I^G(J
^{PC})=2^+(2^{++})$ which should be looked for in the $\rho^\pm\rho^0$ decay 
channels in the reactions $\gamma N\to\rho^\pm\rho^0N$ and $\gamma N\to\rho^\pm
\rho^0\Delta$. The full classification of the $\rho^\pm\rho^0$ states by their
quantum numbers is presented. A simple model for the spin structure of the $
\gamma p\to f_2(1270)p$, $\gamma p\to a^0_2(1320)p$, and $\gamma N\to X^\pm
(N,\,\Delta)$ reaction amplitudes is formulated and the tentative estimates of
the corresponding cross sections at the incident photon energy $E_\gamma\approx
6$ GeV are obtained: $\sigma(\gamma p\to f_2(1270)p)\approx0.12$ $\mu$b, $
\sigma(\gamma p\to a^0_2(1320)p)\approx0.25$ $\mu$b, $\sigma(\gamma N\to X^\pm
N\to\rho^\pm\rho^0N)\approx0.018$ $\mu$b, and $\sigma(\gamma p\to X^-\Delta^{++
}\to\rho^-\rho^0\Delta^{++})\approx0.031$ $\mu$b. The problem of the $X^\pm$
signal extraction from the natural background due to the other $\pi^\pm\pi^0
\pi^+\pi^-$ production channels is discussed. In particular the estimates are
presented for the $\gamma p\to h_1(1170)\pi^+n$, $\gamma p\to\rho'^{\,+}n\to
\pi^+\pi^0\pi^+\pi^-n$, and $\gamma p\to\omega\rho^0p$ reaction cross sections.
Our main conclusion is that the search for the exotic $X^\pm(2^+(2^{++}))$
states is quite feasible at JEFLAB facility. The expected yield of the $\gamma
N\to X^\pm N\to\rho^\pm\rho^0N$ events in a 30-day run at the 100\% detection
efficiency approximates $2.8\times10^6$ events.\\[0.5cm]
PACS number(s): 13.60.Le, 12.40.Nn, 14.40.Ev\\[4.5cm]
email: achasov@math.nsc.ru\\[1pt] email: shestako@math.nsc.ru
\newpage \begin{flushleft} {\bf I.\ \ INTRODUCTION} \end{flushleft}

The intensive photon beam with an energy of 6 GeV and the CLAS detector at
Jefferson Laboratory (JEFLAB) will allow us to perform, to a high accuracy,
measurements of the multi-meson photoproduction processes [1-4]. Hunting for
various unusual resonant states will be one of the main directions of
forthcoming investigations [1-4]. In particular, similar to the current $e^+
e^-$ experiments [5-7], the rare radiative decays $\phi\to f_0(980)\gamma\to\pi
\pi\gamma$ and $\phi\to a^0_0(980)\gamma\to\pi^0\eta\gamma$, which are an
excellent laboratory for studying the makeup of the $f_0(980)$ and $a^0_0(980)$
[8-10], will be looked for in the reaction $\gamma p\to\phi p$. It is also
proposed to study some states from the $J^{PC}=0^{--}$, even$^{+-}$, odd$^{-+}
$ exotic series, for example, the $\hat\rho(1600)$ state with $J^{PC}=1^{-+}$ 
in the reactions $\gamma p\to(\rho\pi,\,\eta\pi,\,\eta'\pi,\,f_1\pi)N$ [2-4].

In the present work, we show that the list of exotic mesons planned for
studying in photoproduction at JEFLAB and other facilities can be extended by 
the states $X^\pm(1600)$ with
$J^{PC}=2^{++}$ being members of an isotopic multiplet with the isospin $I=2$.
These states should be looked for in the $\rho^\pm\rho^0$ decay channels in the
reactions $\gamma p\to\rho^+\rho^0(n,\Delta^0)$, \ $\gamma n\to\rho^-\rho^0(p,
\Delta^+)$, \ $\gamma p\to\rho^-\rho^0\Delta^{++}$, and $\gamma n\to\rho^+\rho
^0\Delta^-$ which can occur via the $\rho$ Regge pole exchange.

As is known, the neutral isotensor tensor state $X^0(1600,\,I^G(J^{PC})=2^+(2^
{++}))$ [11] has been observed near the threshold in the reactions $\gamma
\gamma\to\rho^0\rho^0$ [12,13] and $\gamma\gamma\to\rho^+\rho^-$ [14,15] (see,
for reviews, Refs. [16,17] and the ARGUS data shown in Figure 1, as an
illustration of the situation in the $\gamma\gamma$ collisions). The
specific features due to this state in the reactions $\gamma\gamma\to\rho\rho$
were predicted in Refs. [18,19] on the basis of the $q^2\bar q^2$ MIT-bag model
[20]. While a resonance interpretation of the data on the reactions $\gamma
\gamma\to\rho\rho$ seems to us to be most adequate, it is not yet completely
unquestionable and finally established (see, for example, discussions in Refs.
[16,17,21,22]). Similar to the other candidates in ``certified" exotic states
[23], the state $X^0(1600,\,2^+(2^{++}))$ is in need of further confirmations
and it is not improbable that just photoproduction of its charged partners,
$X^\pm$, will become crucial in this respect \footnote{The cross sections for
hadroproduction of the $X^0(1600,\,2^+(2^{++}))$ doubly charged partners were
estimated in Ref. [24].}.

In Sec. II, we classify the $\rho^\pm\rho^0$ states by their quantum numbers,
indicate those states which can be essential near the nominal $\rho\rho$
threshold, and briefly discuss the resonances known coupled to the $\pi^\pm\pi
^0\pi^+\pi^-$ channels which can be the sources of the background for the $X^
\pm(2^+(2^{++}))$ signals.
In Sec. III, using the available information on the processes $\gamma\gamma
\to f_2(1270)\to\pi\pi$, \ $\gamma\gamma\to a^0_2(1320)\to\pi^+\pi^-\pi^0
,\,\pi^0\eta$, and $\gamma\gamma\to\rho\rho$, the vector dominance model (VDM),
and the factorization property of the Regge pole exchanges, we establish the
spin structure for the $\gamma p\to f_2(1270)p$, $\gamma p\to a^0_2(1320)p$,
and $\gamma N\to X^\pm(N,\,\Delta)$ reaction amplitudes and estimate the values
of the corresponding cross sections. At the incident photon energy $E_\gamma
\approx6$ GeV, we obtain $\sigma(\gamma p\to f_2(1270)p)\approx0.12$ $\mu$b,
$\sigma(\gamma p\to a^0_2(1320)p)\approx0.25$ $\mu$b,
$\sigma(\gamma N\to X^\pm N\to\rho^\pm\rho^0N)\approx0.018$ $\mu$b and
$\sigma(\gamma p\to X^-\Delta^{++}\to\rho^-\rho^0\Delta^{++})=
3\,\sigma(\gamma p\to X^+\Delta^0\to\rho^+\rho^0\Delta^0)\approx0.031$ $\mu$b.

In Sec. IV, we discuss, mainly by the example of the reactions $\gamma N\to\pi
^\pm\pi^0\pi^+\pi^-N$, the problem of the $X^\pm$ state extraction from the
natural background caused by the other $\pi^\pm\pi^0\pi^+\pi^-$ production
channels and estimate a number of relevant partial cross sections.

Using, as a guide, information on the statistics planned for the rare decays of
the $\phi$ meson produced in the reaction $\gamma p\to\phi p$ [1,2], we come
to the conclusion that the search for the exotic $X^\pm(2^+(2^{++}))$ states
near the $\rho\rho$ threshold at JEFLAB facility should be expected quite
successful. New knowledge about hadron spectroscopy, which will be obtained as
a result of such measurements, seem to be extremely important.

\begin{flushleft} {\bf II.\ \ STATES OF THE $\rho^\pm\rho^0$ SYSTEM}
\end{flushleft}
These states have the positive $G$ parity. Their classification by the total
isospin $I$, total moment $J$, \,$P$ parity, $C$ parity of a neutral component
of the isotopic
multiplet, total spin $S$, and total orbital angular moment $L$ is presented in
Table I. As seen from the table, five of eight series of the $\rho^\pm
\rho^0$ states are exotic, i.e. forbidden in the $q\bar q$ system \footnote{
Notice that we do not include in Table I two exotic $\rho^\pm\rho^0$ states
with $I^G(J^{PC})=1^+(0^{--})$ and $1^+(0^{+-})$ because they correspond to the
$^{2S+1}L_J=^3P_0$, and $^1S_0$, $^5D_0$ configurations that are
forbidden by Bose symmetry in the limit of the stable $\rho$ mesons. Therefore,
in the realistic case, the production amplitudes of two unstable $\rho$ mesons
with the four-momenta $q_1$ and $q_2$ have to be proportional, for these
states, to the factor $(q^2_1-
q^2_2)$ which forbids a ``simultaneous" appearance of two $\rho$ peaks in the
$P$-wave mass spectra of the final $(\pi\pi)_1$ and $(\pi\pi)_2$ systems.
For the same reason, we omit some $^{2S+1}L_J$ configurations in the
right column of Table I, for example, the $^3S_1$ configuration for the
$I^G(J^{PC})=2^+(1^{++})$ state and $^3D_2$ for $2^+(2^{++})$.}.
The specific examples of the resonance states exist so far only in the first,
second, and eight series \footnote{The $\rho_2$ state from the third series
with $I^G(J^{PC})=1^+(2^{--})$ and with the expected mass near 1700 MeV [25]
and also its isoscalar partners $\omega_2$ and $\phi_2$ have not yet been
observed in multibody mass spectra [26]. However, the current data on the 
reaction $\pi^-p\to a^0_0(980)n$ are conclusively indicative of the $\rho_2$
Regge pole exchange [27]. Besides the $a_0\pi$ and $\rho\rho$ channels, the 
$\rho_2$ state may be also coupled to $\omega\pi$, $a_2\pi$, $a_1\pi$, $h_1\pi
$, $b_1\eta$, $\rho(\pi\pi)_{S\mbox{-wave}}$, and $\rho f_2$ systems.}. Among
the states with even $J$, only the exotic ones with $I^G(J^{PC})=2^+(0^{++})$
and $2^+(2^{++})$ possess the $^{2S+1}L_J$ configurations with $L=0$ and
therefore can, in principle, effectively manifest themselves near the nominal
$\rho\rho$ threshold ($2m_\rho\approx1540$ MeV). Note that one can speak with
confidence about coupling to the $\rho\rho$ channel only in the case of the $
\rho_3(1690)$ and $X(1600,\,2^+(2^{++}))$ states. Indeed, in the dominant $
\rho_3(1690)\to4\pi$ decay, the $\rho\rho$ mode may reach of 50\% [26], and the
$X^0(1600,\,2^+(2^{++}))$ state has been discovered for the first time just in
the $\rho\rho$ channel (see the Introduction). The $b_1(1235)$ resonance lies
deeply under the $\rho\rho$ threshold and has been observed the in the 
four-pion channel 
only in the $\omega\pi$ mode [26]. As for the $\rho(1700)\to4\pi$ decay
then the available data do not contradict to the absent of the $\rho\rho$
component in this decay [26] \footnote{In this context the notation $\rho(1700)
$ [26] is used somewhat conditionally. Actually, by the $\rho(1700)$ we mean
the whole $4\pi$ enhancement with $I^G(J^{PC})=1^+(1^{--})$ which has been
observed in the reactions $\gamma p\to\pi^+\pi^-\pi^+\pi^-p$ and $\gamma p\to
\pi^+\pi^-\pi^0\pi^0p$ (however, perhaps with the $\rho_3(1690)$ admixture
[28]), i.e. the ``old" $\rho'$ (or $\rho(1600)$) resonance [29]. Taking into 
account the splitting of the $\rho'$ into two $\rho$-like resonances ($\rho(145
0)$ and $\rho(1700)$ [30]) in photoproduction, according to the available data
[26,31-36], is not critical at least for our purposes.}. 
However, for any $4\pi$ decay
mechanisms the $\rho(1700)$ and $\rho_3(1690)$ resonance contributions to the
reactions $\gamma N\to\pi^\pm\pi^0\pi^+\pi^-(N,\Delta)$ should be considered as
a possible important and, may be, major background for signals from the $X^
\pm(1600, 2^+(2^{++}))$ production. Note that the $\rho^\pm(1700)$ and $\rho_3^
\pm(1690)$ states have not yet been observed in photoproduction and that the
available results on diffractive $\rho^0(1700)$ and $\rho_3^0(1690)$
photoproduction for $E_\gamma<10$ GeV are in need of refinements.\newpage

\begin{flushleft} {\bf III.\ \ ESTIMATES OF THE $f_2(1270)$,\ $a^0_2(1320)$,
AND $X^\pm(1600, 2^+(2^{++}))$ PHOTOPRODUCTION CROSS SECTIONS}\end{flushleft}

We shall assume that the $\gamma p\to f_2p\,$, $\gamma p\to a^0_2p$ and $\gamma
N\to X^\pm(N,\,\Delta)$ reaction cross sections at high energies are dominated
by the Regge pole exchanges with natural parity, i.e. by the $\rho^0$,
$\omega$, and $\rho^\pm$ Regge poles \footnote{Wherever possible,
we do not indicate, for short, the resonance masses, i.e., we write, for
example, $f_2$ instead of $f_2(1270)$, etc.}. Note that the one-pion exchange
is forbidden in these reactions and we neglect by the $b_1,\,h_1,\,\rho_2$, and
$\omega_2$ exchanges with unnatural parity. In order to establish the spin
structure for the $\gamma p\to f_2p\,$, $\gamma p\to a^0_2p$, and $\gamma N\to
X^\pm(N,\,\Delta)$ reaction amplitudes in the Regge region and estimate the
corresponding cross sections we shall first discuss the $f_2,\,
a_2^0$, and $X^0$ production in $\gamma\gamma$ collisions (our approach is
closely related essentially with that of Stodolsky and Sakurai [37] to the
construction of the $\rho$ meson exchange model for $\Delta^{++}$ 
production in the reaction $\pi^-p\to\pi^0\Delta^{++}$).

As is known, in the c.m. system in the reactions $\gamma\gamma\to f_2\to\pi\pi$
[38], $\gamma\gamma\to a^0_2\to(\pi^+\pi^-\pi^0,\, \pi^0\eta)$ [39,40] and $
\gamma\gamma\to\rho\rho$ near the threshold [13,15] (see Fig. 1), formation of
the tensor $(J^P=2^+)$ resonances occur mainly in the $J_z=\pm2$ spin states
with the spin quantization $z$-axis taken along the momentum one of two
photons; in so doing $J_z=\lambda_1-\lambda_2$, where $\lambda_1$ and $\lambda
_2$ are the photon helicities. Two-photon excitation of the $2^+$ resonances 
in the pure $J_z=\pm2$ states corresponds to the production amplitude of the 
form $\,g_{2^+\gamma\gamma}\,T^{\lambda*}_{\mu\nu}\,F^{\gamma_1}_{\mu\tau}\,F^{
\gamma_2}_{\nu\tau}$\, [18,41,42], where $T^{\lambda*}_{\mu\nu}$ is the
symmetric traceless polarization tensor for the final $2^+$ resonance with
helicity $\lambda$, $F^{\gamma_i}_{\mu\nu}=k_{i\mu}\epsilon^{\lambda_i}_
\nu(k_i)-k_{i\nu}\epsilon^{\lambda_i}_\mu(k_i)$, and $\epsilon^{\lambda_i}_
\mu(k_i)$ is the usual polarization vector for the photon with the
four-momentum $k_i$ and helicity $\lambda_i=\pm1$\, ($i=1,2$). Indeed, in the
$\gamma\gamma$ c.m. system, the spatial part of the tensor $I_{\mu\nu}=(F^{
\gamma_1}_{\mu\tau}\,F^{\gamma_2}_{\nu\tau}+F^{\gamma_2}_{\mu\tau}\,F^{\gamma_
1}_{\nu\tau})/2$ for $\lambda_1=\lambda_2$ has the form $\,|\vec
k_1|^2[\epsilon^1_l(k_1)\epsilon^1_m(k_2)+\epsilon^{-1}_l(k_1)\epsilon^{-1}_m(
k_2)+\epsilon^0_l(k_1)\epsilon^0_m(k_2)]$ (where $\vec\epsilon\,^0(k_i)=\vec k_
i/|\vec k_i|\,$ and $l,m=1,2,3$) and corresponds to the wave function of the
$\gamma\gamma$ system with $J=0$ which is orthogonal to $T^{\lambda*}_{lm}$
with $\lambda=J_z=0$, whereas, for $\lambda_1=-\lambda_2=\pm1,\,$ $I_{lm}=2|
\vec k_1|^2\epsilon^{\pm1}_l(k_1)\epsilon^{\mp1}_m(k_2)$ corresponds to the
states of the $\gamma\gamma$ system with the total moment $J=2$ and $J_z=\pm2$.

In the spirit of the vector meson dominance (VDM) and naive quark model, we 
shall consider that the
$\gamma V\to2^+$ transition amplitudes have the form $\,g_{2^+\gamma V}\,T^{
\lambda*}_{\mu\nu}\,F^{\gamma}_{\mu\tau}\,F^V_{\nu\tau}\,$ (where $V=\rho,\,
\omega,\,$ $F^V_{\nu\tau}=k_{V\nu}\epsilon^{\lambda_V}_\tau(k_V)-k_{V\tau}
\epsilon^{\lambda_V}_\nu(k_V),$\, and $\,\epsilon^{\lambda_V}\tau(k_V)$ is the
polarization vector of the $V$ meson with the four-momentum $k_V$ and helicity
$\lambda_V$) and that the coupling constants $g_{2^+\gamma V}$ and $g_{2^+
\gamma\gamma}$ obey the following relations: \begin{equation}
g_{f_2\gamma\rho}=3g_{f_2\gamma\omega}=g_{a_2\gamma\omega}=3g_{a_2\gamma\rho}=
\frac{9}{10}g_{f_2\gamma\gamma}\left(\frac{f_\rho}{e}\right)=
\frac{1}{2}g_{a_2\gamma\gamma}\left(\frac{f_\omega}{e}\right)=
\frac{1}{2}g_{a_2\gamma\gamma}\left(\frac{3f_\rho}{e}\right)\ . \end{equation}
Small deviations known from these relations are not of principle for the
following estimates. Using Eq. (1) and the conventional values for
the widths [26] \begin{equation}
\Gamma_{f_2\gamma\gamma}=\frac{g^2_{f_2\gamma\gamma}}{4\pi}\,\frac{m^3_{f_2}}
{80}\approx2.45\,\mbox{keV\ \ \ and\ \ \ }\Gamma_{\rho^0e^+e^-
}\approx\frac{\alpha^2}{3}\,\frac{m_\rho}{(f^2_\rho/4\pi)}=6.77\,\mbox{keV}\ ,
\end{equation} where $\alpha=e^2/4\pi=1/137$, we find that $f^2_\rho/4\pi
\approx2.02$, \begin{equation}\frac{g^2_{f_2\gamma\rho}}{4\pi}\approx0.0212\,
\mbox{GeV}^{-2}\,,\ \ \ \mbox{and}\ \ \ \Gamma_{f_2\gamma\rho}=\frac{g^2_{f_2
\gamma\rho}}{4\pi}\,\frac{|\vec k_\gamma|^3}{5}\,\left[1+\frac{1}{2}\,r+
\frac{1}{6}\,r^2\right]\approx340\,\mbox{keV}\,,\end{equation} where $r=m^2_
\rho/m^2_{f_2}\,$ and $\,|\vec k_\gamma|=m_{f_2}(1-r)/2$.

Let us now construct the $s$-channel helicity amplitudes $M^{(\rho)}_{\lambda_
{f_2}\lambda'_p\lambda_\gamma\lambda_p}$ for the reaction $\gamma p\to f_2p$
corresponding to the elementary $\rho$ exchange. As is well known, at high
energies and fixed momentum transfers, such amplitudes possess one of the major
properties
of the Regge pole amplitudes, namely, the factorization property of the spin
structures for mesonic and baryonic vertices. Thus, in the c.m. system,
\begin{equation} M^{(\rho)}_{\lambda_{f_2}\lambda'_p\lambda_\gamma\lambda_p}=
V^{(\rho)}_{\lambda_{f_2}\lambda_\gamma}(t)\left(\frac{-2s}{t-m^2_\rho}\right)
V^{(\rho)}_{\lambda'_p\lambda_p}(t)\ ,\end{equation} where $s=(k+p)^2$, $t=(q-k
)^2$, $k+p=q+p'$,\, $k,q,p,p'$ and $\lambda_\gamma,\lambda_{f_2},\lambda_p,
\lambda_p'$ are the four-momenta and helicities of the photon, $f_2$ meson,
initial and final protons respectively. According to the model suggested above
for the $\gamma\rho f_2$ interaction, the corresponding vertex function looks 
as follows: \begin{eqnarray}V^{(\rho)}_{\lambda_{f_2}\lambda_\gamma}(t)=-(g_{
f_2\gamma\rho}/2)\,\xi^{\lambda_{f_2}*}_{jl}\times \qquad\qquad\qquad\qquad
\nonumber\\\times\left[\epsilon^{\lambda_\gamma}_j\Delta_l
+\frac{(\vec\epsilon^{\,\lambda_\gamma}\vec\Delta)}{m_{f_2}}n_{0j}\Delta_l
+\frac{(\vec\epsilon^{\,\lambda_\gamma}\vec n)}{m_{f_2}}n_{j}n_{0l}
+\frac{1}{2}(\vec\epsilon^{\,\lambda_\gamma}\vec\Delta)n_{0j}n_{0l}
\left(1-\frac{t}{m^2_{f_2}}\right)\right]\ , \end{eqnarray}
where $\xi^{\lambda_{f_2}*}_{jl}$ is the three-dimensional tensor spin wave
function of the final $f_2$ meson in its rest frame ($j,l=1,2,3$), $\vec\Delta=
\vec q-\vec k$, $t\approx-\vec\Delta^2$, $\vec\epsilon^{\,\lambda_\gamma}$ is
the three-dimensional polarization vector of the photon, the vector $\vec n_0=
\vec k/|\vec k|$ is aligned along the $z$ (or 3) axis, and the vector $\vec n=
[\vec n_0\,\vec\Delta]$ is aligned along the $y$ (or 2) axis which is the
normal to the reaction plane. Note that the explicit form of the vertex
function $V^{(\rho)}_{\lambda'_p\lambda_p}(t)$ will not be required in the
following; here we also neglect $t_{\mbox{min}}\approx-m^2_pm^4_{f_2}/s^2$. 
Eq. (5) yields \begin{equation} V^{(\rho)}_{\pm2\pm1}(t)=
\pm\,\frac{g_{f_2\gamma\rho}\sqrt{-t}}{2\sqrt{2}}\,,\ \ \ \ \ V^{(\rho)}_
{\pm1\pm1}(t)=\frac{g_{f_2\gamma\rho}\,t}{2\sqrt{2}\,m_{f_2}}\,,\ \ \ \ \
V^{(\rho)}_{0\pm1}(t)=\mp\,\frac{g_{f_2\gamma\rho}\,t\,\sqrt{-t}}{4\sqrt{3}\,
m^2_{f_2}}\,. \end{equation} Moreover, in our model, the relative contributions
of the helicity amplitudes with $\lambda_{f_2}-\lambda_\gamma=\pm3$ and $\pm2$
vanish asymptotically. Eq. (6) implies that, for $-t<1$ GeV$^2$, the
contributions to the differential cross sections from the amplitudes with
$\lambda_{f_2}=\pm1$ and $\lambda_{f_2}=0$ are suppressed relative to those of
the amplitudes with $\lambda_{f_2}=\pm2$ by the factors $-t/m^2_{f_2}$ and $t^
2/6m^4_{f_2}$ respectively. Thus, our model predicts a dominance of the $f_2$
production amplitudes with $\lambda_{f_2}=\pm2$ in the region $-t<1$ GeV$^2$.
Going to the real physical amplitudes caused by the $\rho$ Regge pole exchange,
$\widehat{M}^{(\rho)}_{\lambda_{f_2}\lambda'_p\lambda_\gamma\lambda_p}$, we
shall accept this prediction as a natural assumption and take into account
hereinafter just the amplitudes $\widehat{M}^{(\rho)}_{\pm2\lambda'_p\pm1
\lambda_p}$. Denote the Regge vertex functions (Regge residues),
which appear in
$\widehat{M}^{(\rho)}_{\lambda_{f_2}\lambda'_p\lambda_\gamma\lambda_p}$, by $
\widehat{V}^{(\rho)}_{\lambda_{f_2}\lambda_\gamma}(t)$ and $\widehat{V}^{(\rho)
}_{\lambda'_p\lambda_p}(t)$. The parity conservation and residue factorization
[43] yield $\widehat{M}^{(\rho)}_{2\lambda'_p1\lambda_p}=-\widehat{M}^{(
\rho)}_{-2\lambda'_p-1\lambda_p}=(-1)^{\lambda'_p-\lambda_p}\widehat{M}^{(\rho)
}_{2-\lambda'_p1-\lambda_p}$. Therefore we deal only with two independent
amplitudes, for example, $\widehat{M}^{(\rho
)}_{2\frac{1}{2}1\frac{1}{2}}$ and $\widehat{M}^{(\rho)}_{2\frac{1}{2}1-\frac{1}{
2}}$. Note that the $\omega$ exchange in the reaction $\gamma p\to f_2p$
quadruplicates the $\widehat{M}^{(\rho)}_{2\frac{1}{2}1\frac{1}{2}}$
contribution to the cross section. Indeed, assuming, as in the naive quark
model, that  $\widehat{V}^{(\omega)}_{\lambda_{f_2}\lambda_\gamma}(t)=\widehat
{V}^{(\rho)}_{\lambda_{f_2}\lambda_\gamma}(t)/3$ (see also Eq. (1)) and $
\widehat{V}^{(\omega)}_{\lambda'_p=\frac{1}{2},\lambda_p=\frac{1}{2}}(t)=3
\widehat{V}^{(\rho)}_{\lambda'_p=\frac{1}{2},\lambda_p=\frac{1}{2}}(t)$, and
also degeneracy of the $\rho$ and $\omega$ Regge trajectories, $\alpha_\omega(t)=
\alpha_\rho(t)$, we obtain $\widehat{M}^{(\omega)}_{2\frac{1}{2}1\frac{1}{2}}=
\widehat{M}^{(\rho)}_{2\frac{1}{2}1\frac{1}{2}}$. On the other hand, the $
\omega$ exchange amplitude with a helicity flip in the nucleon vertex is
assumed negligible because $\widehat{V}^{(\omega)}_{\frac{1}{2}-\frac{1}{2}}(t)
<<\widehat{V}^{(\rho)}_{\frac{1}{2}-\frac{1}{2}}(t)$ and in addition $\widehat{
V}^{(\omega)}_{\frac{1}{2}-\frac{1}{2}}(t)/\sqrt{-t/1\mbox{GeV}^2}<<\widehat{V}
^{(\omega)}_{\frac{1}{2}\frac{1}{2}}(t)$ (see, for example [44-47]). Finally,
for the reaction $\gamma p\to f_2p$ in the standard normalization we have
\begin{eqnarray} \sigma(\gamma p\to f_2p)=\frac{1}{16\pi s^2}\,\int\,\left[\,4
\,\left|\widehat{M}^{(\rho)}_{2\frac{1}{2}1\frac{1}{2}}\right|^2+\left|\widehat
{M}^{(\rho)}_{2\frac{1}{2}1-\frac{1}{2}}\right|^2\,\right]dt\,=\,4\,\sigma^{(
\rho)}_{nf}\left(1+\frac{1}{4}\,R\right)\,,\end{eqnarray} where the integral is
taken over the region $0<-t<1$ GeV$^2$ which gives the main contribution to the
cross section, $\sigma^{(\rho)}_{nf}$ denotes the cross section caused by the
$\rho$ exchange amplitude $\widehat{M}^{(\rho)}_{2\frac{1}{2}1\frac{1}{2}}$
without the nucleon helicity flip, and $R=\sigma^{(\rho)}_f/\sigma^{(\rho)}_{nf
}$ is the ratio of the cross section $\sigma^{(\rho)}_f$ caused by the
amplitude $\widehat{M}^{(\rho)}_{2\frac{1}{2}1-\frac{1}{2}}$ with the nucleon
helicity flip to $\sigma^{(\rho)}_{nf}$. To estimate $R$ we use the data on the
$\pi^-p\to\pi^0n$ reaction differential cross section which are described
remarkably well in terms of the $\rho$ Regge pole exchange [48]. Assuming the
approximate equality of the slopes $\Lambda$ for the Regge amplitudes in
question
\footnote{Here we have in mind the usual exponential parametrization, according
to which any Regge amplitude is taken to be proportional to $e^{\Lambda t}$,
where the slope $\Lambda=\Lambda^0+\alpha'\ln(s/s_0)$,\, $\alpha'$ is that of
the Regge pole trajectory, $s_0=1$ GeV$^2$, and $\Lambda^0$ is determined by
fitting to the data.}, together with factorization of the Regge pole residues,
and using the results of Ref. [48], we put \begin{eqnarray}
R\approx\sigma^{(\rho)}_f(\pi^-p\to\pi^0n)/\sigma^{(\rho)}_{nf}(\pi^-p\to
\pi^0n)\approx1.5\,.\end{eqnarray} Note that this estimate for $R$ can
be considered as a lower bound. The point is that $R$ is proportional to $1/2
\Lambda$ and for the $\pi^-p$ charge exchange at 6 GeV [48] $2\Lambda\approx9$
GeV$^{-2}$ which, generally speaking, is larger than for many other similar
reactions.

Let us now obtain a representation similar to Eq. (7) for the cross section of
the reaction $\gamma p\to a^0_2p$. According to the naive quark counting rules,
any $\rho$ exchange amplitude for $\gamma p\to a^0_2p$ is three times smaller
than the corresponding one for $\gamma p\to f_2p$ and the $\omega$ exchange
amplitude without the proton helicity flip for $\gamma p\to a^0_2p$ is three 
times larger than that for $\gamma p\to f_2p$ (see also Eq. (1)); i.e., the 
reaction $\gamma p\to a^0_2p$ is dominated by the $\omega$ exchange. Thus, in 
terms of the $\rho$ exchange amplitudes pertaining to the reaction $\gamma p
\to f_2p$, for which $\widehat{M}^{(\omega)}_{2\frac{1}{2}1\frac{1}{2}}=
\widehat{M}^{(\rho)}_{2\frac{1}{2}1\frac{1}{2}}$ (see also Eqs. (7),(8)), the 
cross section for $\gamma p\to a^0_2p$ is given by
\begin{eqnarray} \sigma(\gamma p\to a^0_2p)=\frac{1}{16\pi s^2}\,\int\,\left[\,
\frac{100}{9}\,\left|\widehat{M}^{(\rho)}_{2\frac{1}{2}1\frac{1}{2}}\right|^2+
\frac{1}{9}\,\left|\widehat{M}^{(\rho)}_{2\frac{1}{2}1-\frac{1}{2}}\right|^2\,
\right]dt\,=\nonumber\\[0.1cm] =\,\frac{100}{9}\,\sigma^{(\rho)}_{nf}\left(1
+\frac{1}{100}\,R\right)\approx\,\frac{100}{9}\,\sigma^{(\rho)}_{nf}\,.\qquad
\qquad\ \ \ \ \ \end{eqnarray} Further, let us notice that exactly the same
relation between the $\omega$ and $\rho$ exchanges, as in $\gamma p\to a^0_2p$,
takes place in the reaction $\gamma p\to\pi^0p$ the cross section of which is
dominated by the same exchanges [45-47,49-51]. Furthermore, the helicity change
in the mesonic Regge vertices is equal to 1 both in $\gamma p\to\pi^0p$ and in
our model for the reaction $\gamma p\to a^0_2p$ (and $\gamma p\to f_2p$),
so that the all corresponding residues are proportional to $\sqrt{-t}$.
Defining the $\omega\to\pi^0\gamma$ decay amplitude in the conventional form
$g_{\omega\gamma\pi}\,\varepsilon_{\mu\nu\tau\kappa}\epsilon^{\lambda_\omega
}_\mu(k_\omega)k_{\omega\nu}\epsilon^{\lambda_\gamma}_\tau(k_\gamma)k_{\gamma
\kappa}$, we find in the case of the elementary $\omega$ and $\rho$ exchanges
that without any numerical factors $\sigma(\gamma p\to a^0_2p)/\sigma(\gamma
p\to\pi^0p)=g^2_{a_2\gamma\omega}/g^2_{\omega\gamma\pi}$. Thus, it is
reasonable to suppose that, in the realistic case of the Reggeized $\omega$ and
$\rho$ exchanges, such a ratio can be estimated as follows: \begin{equation}
\frac{\sigma(\gamma p\to a^0_2p)}{\sigma(\gamma p\to\pi^0p)}\approx\frac{g^2_{
a_2\gamma\omega}}{g^2_{\omega\gamma\pi}}\,\frac{\Lambda^2_\pi}{\Lambda^2_{a_2}
}\,, \end{equation} where $\Lambda_\pi$ and $\Lambda_{a_2}$ are the Regge
slopes for the $\pi^0$ and $a_2^0$ photoproduction amplitudes respectively.
To get a feeling for the influence of the slopes in the absence of any
information obout $\Lambda_{a_2}$, we make the {\it ad hoc} assumption
that $\Lambda_{a_2}\approx\Lambda_\pi/1.225$, or $\Lambda^2_\pi/\Lambda^2_{a_2}
\approx1.5$. According to Refs. [49-51], $\sigma(\gamma p\to\pi^0p)\approx0.32$
$\mu$b at $E_\gamma\approx6$ GeV. From the relations $\Gamma_{\omega\gamma\pi}=
(g^2_{\omega\gamma\pi}/4\pi)|\vec k_\gamma|^3/3\approx715$ keV [26] and Eqs.
(1), (3) we get $g^2_{\omega\gamma\pi}/4\pi\approx0.0394$ GeV$^{-2}$ and
$g^2_{a_2\gamma\omega}/g^2_{\omega\gamma\pi}\approx0.538$ which need be
substitute in Eq. (10). Putting all this together, we find from Eqs. $(7)-(10)$
that at $E_\gamma\approx6$ GeV we can expect \begin{equation}
\sigma(\gamma p\to f_2p)\approx0.12\ \mu\mbox{b}\qquad\mbox{and}
\qquad\sigma(\gamma p\to a^0_2p)\approx0.25\ \mu\mbox{b}\,. \end{equation}
Unfortunately, the data on the reactions $\gamma p\to f_2p$ and $\gamma p\to
a^0_2p$ are very poor. From the experiments as carried out at comparable
energies in the late 60s it is know only that $\sigma(\gamma p\to f_2
p\to\pi\pi p)=0.06\pm0.30\ \mu$b for $4.5<E_\gamma<5.8$ GeV [52],
$\sigma(\gamma p\to a^0_2p)<0.35\ \mu$b for $2.2<E_\gamma<5.8$ GeV
[52], $\sigma(\gamma p\to f_2p\to\pi\pi p)<0.5\ \mu$b and $\sigma(\gamma p
\to a^0_2p\to\pi^+\pi^-\pi^0 p)<0.4\ \mu$b at 5.25 GeV [53], and also
$\sigma(\gamma p\to f_2p\to\pi^+\pi^-p)\leq0.7\pm0.4\ \mu$b at 4.3 GeV [54].
   At higher energies, the reactions $\gamma p\to\pi^+\pi^-p$
   and $\gamma p\to\pi^+\pi^-\pi^0p$ are dominated by two- and
   three-pion states producing mainly via the Pomeron exchange. Such states
   have been investigated in some detail. However, the reached accuracy do
   not yet allow certain conclusions to be made concerning the presence of
   the $C$-odd Regge exchanges. Let us consider as an example the data on the
   reaction $\gamma p\to\pi^+\pi^-\pi^0p$ in the energy range $20-70$ GeV
   obtained by the Omega Photon Collaboration [53]. Assuming the $1/E_\gamma$
   energy dependence for $\sigma(\gamma p\to a^0_2p)$, which certainly gives
   an upper limit of the cross section as $E_\gamma$ increases, and thus
   extrapolating the value for $\sigma(\gamma p\to a^0_2p)$ from Eq. (11) to
   the region $20<E_\gamma<70$ GeV, we obtain the averaged $a^0_2$ production
   cross section times branching ratio of $a_2$ to $\rho\pi$ of about 24 nb.
   Note that the observed three-pion production cross section in the three-pion
   mass range from 1.2 to 1.5 GeV is nearly 600 nb [53]. Furthermore, one can
   see that the height of the $a^0_2$ peak above a large and smooth background
   in the three-pion mass spectrum is about a factor of 2.5 smaller than that
   of the observed ``$\omega'(1670)$" peak, because for the latter $\sigma
   \times B\approx100$ nb and the peak width $\approx160$ MeV [53]. Such a
   maximum possible enhancement of the three-pion mass spectrum in the $a^0_2(
   1320)$ region turns out to be somewhat less in magnitude than the existing
   $(8-9)$\% double statistical errors. Thus, it is clear that more efforts
   are needed to provide a reliable observation of the reactions $\gamma p\to
   f_2p$ and $\gamma p\to a^0_2p$. In this connection we would like especially
   to note the reactions $\gamma p\to\pi^0\pi^0 p$ and $\gamma p\to\pi^0\eta p
   $ with the peripherally produced $\pi^0\pi^0$ and $\pi^0\eta$ pairs, which
   can only proceed via C-odd exchanges and therefore have to be
   dominated by the production of the $f_0(980)$, $f_2(1270)$ and $a^0_0(980)
   $, $a^0_2(1320)$ resonances with small background.

Exactly the same approach to the reactions $\gamma N\to X^\pm N\to\rho^\pm\rho^
0N$ leads to the estimate \begin{eqnarray} \sigma(\gamma p\to X^+n
\to\rho^+\rho^0n)=\sigma(\gamma n\to X^-p\to\rho^-\rho^0p)\approx\ \ \ \ \ \
\nonumber\\[0.1cm]\approx\frac{9}{50}\,\sigma(\gamma p\to a^0_2p)\,(1+R)\,\frac
{g^2_{X^\pm\gamma\rho^\pm}\,B(X^\pm\to\rho^\pm\rho^0)}{g^2_{f_2\gamma\rho}}\,
\approx0.018\ \mu\mbox{b}\,.\end{eqnarray} Here we have used Eqs. (3), (8),
(11) and, to avoid the addition model dependent assumptions, estimated the
value of $(g^2_{X^\pm\gamma\rho^\pm}/4\pi)\,B(X^\pm\to\rho^\pm\rho^0)$ using
the data on $\sigma(\gamma\gamma\to\rho^0\rho^0)$ [13] shown in Fig. 1 and the
following transparent relations:\begin{eqnarray}\frac{g^2_{X^\pm\gamma\rho^\pm}
}{4\pi}\,B(X^\pm\to\rho^\pm\rho^0)=\frac{9}{8}\,\frac{g^2_{X^0\gamma\rho^0}}{4
\pi}\,B(X^0\to\rho^0\rho^0)=\frac{9}{8}\,\left(\frac{f_\rho}{e}\right)^2\,
\frac{g^2_{X^0\gamma\gamma}}{4\pi}\,B(X^0\to\rho^0\rho^0)\nonumber\\ \approx
\frac{9}{8}\,\left(\frac{f_\rho}{e}\right)^2\frac{4}{\pi^2\bar m}\,\left(
\frac{1}{2}\,\int \limits_{1.2\,\mbox{\scriptsize GeV}}^{2.2\,\mbox{\scriptsize
GeV}}\,\sigma(\gamma\gamma\to\rho^0\rho^0)\,dW_{\gamma\gamma}\right)\approx0.00
336\ \mbox{GeV}^{-2}\,,\ \ \ \ \ \ \ \ \ \ \end{eqnarray} where $\bar m\approx1
.6$ GeV is the average mass of the enhancement observed in $\gamma\gamma\to\rho
^0\rho^0$, the integral of the cross section $\approx33.2$ nb\,GeV, and we have
put, on the experience of the previous analyses [17,18], that approximately one
half of this value is due to the $X^0$ resonance contribution.

For the reactions $\gamma N\to X^\pm\Delta\to\rho^\pm\rho^0\Delta$ we
also expect \begin{eqnarray}\sigma(\gamma p\to X^-\Delta^{++}\to\rho^-\rho^0
\Delta^{++})=\sigma(\gamma n\to X^+\Delta^-\to\rho^+\rho^0\Delta^-)=\ \ \ \ \ \
\ \nonumber\\[0.1cm]=3\,\sigma(\gamma p\to X^+\Delta^0\to\rho^+\rho^0\Delta^0)=
3\,\sigma(\gamma n\to X^-\Delta^+\to\rho^-\rho^0\Delta^+)\approx0.031\
\mu\mbox{b}\,.\end{eqnarray} This estimate has been obtained simply by
multiplying of that from Eq. (12) by the coefficient 1.75. Here we proceed from
the fact that in the energy region around 6 GeV the cross section for $\pi^+p
\to\pi^0\Delta^{++}$, which is dominated by the $\rho$ Regge pole exchange
[55,56], is approximately $1.5-2$ times than that for the reaction $\pi^-p\to
\pi^0n$ with just the same mechanism [48,56]. Note that approximately the same
ratio takes place for the $\gamma p\to\rho^-\Delta^{++}$ and $\gamma N\to\rho^
\pm N$ reaction cross sections (less the small one-pion exchange contributions)
[57,58].

Let us make two short remarks on the above estimates. Firstly, we consider that
these estimates are rather conservative. Secondly, of course, there are
absorption corrections to the Regge pole models (see, for example, Refs.
[45,46]). They lead to some well known modifications of the $t$ distributions
due to the pure Regge pole exchanges. However, it is reasonable that the $t$
distributions for the reactions having identical Regge pole mechanisms and an
identical spin structure for the dominant helicity amplitudes can remain more
or less similar in shape even in the presence of the absorption contributions.
Therefore, for example, Eq. (10) may well be more general then its derivation
within the framework of the Regge pole approximation. In fact, Eq. (10) can be
modified only by the difference between the absorption corrections to the
$a^0_2$ and $\pi^0$ production amplitudes and so a marked change of the
estimates for the integrated cross sections seems to be unlikely.

At JEFLAB facility, a 6 GeV photon beam, with intensity $5\times10^7$ $\gamma
$'s/sec, will yield about 30 $\phi$'s/sec via the photoproduction process $
\gamma p\to\phi p$, i.e. there can be accumulated about $77.8\times10^6$ $
\gamma p\to\phi p$ events in a 30-day run [1,2]. The cross section for $\gamma
p\to\phi p$ is approximately 0.5 $\mu$b at $E_\gamma\approx6$ GeV [1,2]. Then,
according to our estimate (see Eq. (12)), for the reaction $\gamma N\to X^\pm
N\to\rho^\pm\rho^0N\to\pi^\pm\pi^0\pi^+\pi^-N$ one can expect about $2.8\times1
0^6$ events at the same time. It must be born in
   mind that such a number of events can be accumulated only at the 100\%
   detection efficiency which is certainly inaccessible in practice. If one
   assumes that a photon flux of around $10^7$ $\gamma$'s/sec and a detection
   efficiency of around 10\% are closer to the reality, then one can expect
   approximately 56000 useful events.
For comparison, full statistics collected by the TASSO, CELLO, TPC/2$\gamma$,
PLUTO, and ARGUS groups for the reaction $\gamma\gamma\to\pi^+\pi^-\pi^+\pi^-$
includes 15242 events [17]. At JEFLAB facility, it is planned to obtain several
thousands, tens of thousands, and hundreds of thousands of events for the $\phi
$ meson decays with $B\approx10^{-4}-10^{-2}$ [1,2]. On this scale, the cross
section values indicated in Eqs. (12) and (14) are large and the relevant 
expected significant statistics has not to be wasted. However, there is
also a very important and rather complicated question related to the extraction
of the exotic $X^\pm$ signals from a ``sea" of all possible $\pi^\pm\pi^0\pi^+
\pi^-$ events. It is this question that we want to discuss in the following.

\begin{flushleft} {\bf IV.\ \ ON THE EXTRACTION OF THE $X^\pm(1600,2^+(2^{++}))
$ SIGNALS}\end{flushleft}

Let us consider the reactions $\gamma N\to\pi^\pm\pi^0\pi^+\pi^-N$. In the
first place, it is necessary to obtain some general idea about the most
important channels of these reactions, in particular, about the values of the
corresponding partial cross sections. It is useful to review briefly the
results available for the related and more studied reactions $\gamma N\to\pi^+
\pi^-\pi^+\pi^-N$ in the energy region from 4 to 10 GeV [31-34,52-54,59-69]
\footnote{Here we omit not very essential details concerning the
$\gamma p\to\pi^+\pi^-\pi^+\pi^-p$ and $\gamma n\to\pi^+\pi^-\pi^+\pi^-n$ cross
section difference (in this connection, see, for example, Refs. [61,63,68]).}.
So, the values of the $\gamma N\to\pi^+\pi^-\pi^+\pi^-N$ cross sections lie in
the band from 4 to 7 $\mu$b [67,68]. The reaction $\gamma p\to\pi^+\pi^-\pi^+
\pi^-p$ is strongly dominated by $\rho^0$ and $\Delta^{++}$ production. There
are three the most significant channels in this reaction: $\gamma p\to\pi^+\pi
^-\pi^-\Delta^{++}$, $\gamma p\to\rho^0\pi^-\Delta^{++}$ and $\gamma p\to\rho^0
\pi^+\pi^-p$ [31-33,53,54,60,67] (a similar situation also takes place in 
$\gamma n\to\pi^+\pi^-\pi^+\pi^-n$ [62]). The channels involving $\Delta^{++}$
production define from 30 to 50\% of all $\gamma p\to\pi^+\pi^-\pi^+\pi^-p$
reaction events. Most of the remainder
events are due to $\rho^0$ production. The $\gamma p\to\rho^0\pi^+\pi^-p$
channel is dominated by diffractive $\rho'^{\,0}$ production (see footnote
4). Within the experimental uncertainties, the cross section for $\gamma
p\to\rho'^{\,0}p\to\rho^0\pi^+\pi^-p$ is energy independent and is roughly $1-
2$ $\mu$b [31-33]. Note that the total number of events collected in all
experiments on the reaction $\gamma p\to\pi^+\pi^-\pi^+\pi^-p$ for $4<E_\gamma
<10$ GeV do not exceed $10^4$. Specific methods used for separating a large
number of particular channels in the reaction $\gamma p\to4\pi p$ have been
described in detail in Refs. [31-36,67].

Table II shows the available data on the reactions $\gamma N\to\pi^\pm\pi^0
\pi^+\pi^-N$ for the average incident photon energies from 3.9 to 8.9 GeV.
The total cross sections for $\gamma n\to\pi^-\pi^0\pi^+\pi^-p$ was measured in
four experiments [61-63,68]. In one of them [62], with a total of 151 events,
the cross sections for $\omega$, $\rho^\pm$, $\rho^0$, and $\Delta^0$
production were roughly determined (see three left columns in Table II). In
addition, the channel cross sections for $\omega$ production were measured in
three more experiments [70-72].

Let us now turn to the phenomenological estimates. For definiteness we shall
consider the reaction $\gamma p\to\pi^+\pi^0\pi^+\pi^-n$ and its particular
channels at $E_\gamma\approx6$ GeV (however, the results will also be valid for
$\gamma n\to\pi^-\pi^0\pi^+\pi^-p$). If needed, we shall extrapolate the known
cross section values to $E_\gamma\approx6$ GeV assuming roughly $\sigma\sim E_
\gamma^{-n}$ with $n=2$ and 1 for the one-pion exchange (OPE) mechanism and for
the $\rho$, $a_2$, and $\omega$ exchange ones respectively.

We begin with the channel $\gamma p\to\omega\Delta^+\to\omega\pi^+n$. Using the
data compilations [56,73] and assuming the OPE mechanism dominance, the
factorization property of Regge pole residues, and the approximate equality of 
the slopes for the Regge amplitudes, we get the following estimate
\begin{eqnarray}\sigma(\gamma p\to\omega\Delta^+\to\omega\pi^+n)\approx
\qquad\qquad\qquad\qquad\qquad\nonumber\\[0.1cm]\approx\sigma^{(OPE)}
(\gamma p\to\omega p)\,\frac{4}{9}\,\frac{\sigma(\pi^+p\to\rho^0\Delta^{++})}
{\sigma(\pi^-p\to\rho^0n)}\approx(0.6\,\mu\mbox{b})\,\frac{4}{9}\,2
\approx0.53\,\mu\mbox{b}\,,\end{eqnarray} which is in close agreement with that
of $\sigma(\gamma p\to\omega\Delta^+\to\omega\pi^+n)\approx(0.83\pm0.10\,\mu$b)
$\times(4.5/6)^2\approx0.47\pm0.06\ \mu$b found from the data of Ref. [71]
presented in Table II. If one suppose that not only $\omega\Delta^+$ production
but the whole channel $\gamma p\to\omega\pi^+n$ can also be dominated by $\pi$
exchange between the $\gamma\omega$ and $p\pi^+n$ vertices (certainly, in this
case, the exchanges with the vacuum quantum numbers are also possible), then,
using the data on $\gamma n\to\omega\pi^-p$ [62] presented in Table II, one
obtains $\sigma(\gamma p\to\omega\pi^+n)\approx(1.4\pm0.5\,\mu$b$)\times(4.3/6
)^2\approx0.72\pm0.26\ \mu$b. Owing to the $\omega$ resonance narrowness,
the $\gamma p\to\omega\pi^+n$ channel can be easily separated by
cutting a suitable window in the $\pi^+\pi^-\pi^0$ invariant mass spectrum.

It is interesting that the cross section for peripheral production of the $C
$-odd $\pi^+\pi^-\pi^0$ system with $J^P=1^+$ in $\gamma p\to\pi^+\pi^0\pi^+\pi
^-n$ can be even larger than that for $\omega$ production owing to the $h_1(11
70)$ resonance contribution. Because the $h_1(1170)$ decays mainly into $\rho
\pi$ [26] and its production in $\gamma p\to\pi^+\pi^0\pi^+\pi^-n$ can occur
via one-pion exchange, we can write \begin{eqnarray}\sigma(\gamma p\to h_1\pi^+
n)\approx\sigma(\gamma p\to\omega\pi^+n)\frac{\Gamma_{h_1\gamma\pi}}{\Gamma_{
\omega\gamma\pi}}\approx[(0.53-0.72)\,\mu\mbox{b}]\,2.79\approx(1.48-2)\,\mu
\mbox{b}\,.\end{eqnarray} Here we have used the above estimates for $\sigma(
\gamma p\to\omega\pi^+n)$ and the relation $\Gamma_{h_1\gamma\pi}\approx9\,
\Gamma_{b_1\gamma\pi}\approx9\times0.23\,$MeV$\,\approx2\,$MeV [26] which is
true for ``ideal" octet-singlet mixing in the $J^{PC}=1^{+-}$ nonet. There are
three intermediate states $\rho^+\pi^-$, $\rho^-\pi^+$, and $\rho^0\pi^0$ in
the $h_1\to\rho\pi\to3\pi$ decay. Each of them, combined with the corresponding
kinematical reflections, gives a third of the rate $h_1\to3\pi$. Thus, $\sigma(
\gamma p\to h_1\pi^+n\to\rho^-\pi^+\pi^+n)\approx(0.49-0.67)\,\mu$b, which is
roughly compatible with the value of $\sigma(\gamma n\to\rho^+\pi^-\pi^-p)
\approx0.5\pm0.5\,\mu$b at 4.3 GeV given in Table II. It is obvious that the
channel $\gamma p\to X^+n\to\rho^+\rho^0p$ does not involve $\rho^-$-like
events. Therefore, a careful examination of the $\gamma p\to\rho^-\pi^+\pi^+n$
channel will, probably, permit both peripheral $\rho^-\pi^+$ production and
related $\rho^+\pi^-$ and $\rho^0\pi^0$ events to be successfully selected and
excluded. Together with the contribution from the $h_1(1170)$, there may also
exist the contributions from the $a_1^0$, $a_2^0$, $\pi_2^0$, and $\pi^0(1300)$
resonances in the $\rho^\mp\pi^\pm$-like events.
However, an analysis shows that the cross sections for peripheral
production of these $C$-even resonances via $\rho$ and $\omega$ exchanges in
$\gamma p\to\pi^+\pi^0\pi^+\pi^-n$ should be expected to be small. So, as is
seen from Eqs. (16) and (15), the $\rho^\mp\pi^\pm\pi^+n$, $\rho^0\pi^0\pi^+n$
and $\omega\pi^+n$ production channels can contribute to the $\gamma p\to\pi^+
\pi^0\pi^+\pi^-n$ reaction cross section of about $2-2.5\,\mu$b.

Above we have discussed the peripheral production of the neutral three-pion
systems. Now we consider the processes of peripheral production of the $
\pi^+\pi^-\pi^+$ system, in which the $a_1^+$, $a_2^+$, $\pi_2^+$ and $\pi^+(13
00)$ resonances can manifest themselves, and, in particular, the reaction $
\gamma p\to\pi^+\pi^-\pi^+\Delta^0\to\pi^+\pi^-\pi^+\pi^0n$. Using the above
mentioned data on the reaction $\gamma p\to\pi^+\pi^-\pi^-\Delta^{++}$ for $E_
\gamma<10$ GeV and assuming the peripheral character of $\Delta^{++}$
formation, one can obtain the following tentative estimate:
\begin{equation}\sigma(\gamma p\to\pi^+\pi^-\pi^+\Delta^0\to\pi^+\pi^-\pi^+
\pi^0n)\approx\frac{2}{9}\sigma(\gamma p\to\pi^+\pi^-\pi^-\Delta^{++})\approx
(0.37-0.61)\,\mu\mbox{b}\,.\end{equation} It is clear that such a type of the
$\pi^+\pi^-\pi^+\pi^0n$ events can be
separated, at least, by the specific signs of the $\Delta^0$ resonance.
Formation of $\Delta^+$ in $\gamma p$ collisions accompanied by
$\pi^+\pi^-\pi^0$ production we have already discussed ( the cross section
value for the related process $\gamma n\to\pi^+\pi^-\pi^0\Delta^0$ is presented
in Table II). The final state $\pi^+\pi^0\pi^+\Delta^-$ would also require a
careful study. If there exists the mode $\rho^+\pi^+\Delta^-$ in the channel
$\pi^+\pi^0\pi^-\Delta^-$, then it may be responsible for a possible excess of
the yield of the $\rho^+$ over that of the $\rho^-$ in $\gamma p\to\pi^+\pi^0
\pi^+\pi^-n$. However, the cross section for the $\gamma p\to\rho^+\pi^+
\Delta^-$ channel is difficult to estimate. A similar statement is, 
unfortunately, also true both for $a^+_2\pi^0n$ production with the $\pi^0n$ 
system in the $I=1/2$ state and for the $\gamma p\to\rho^0\pi^+\pi^0n$ channel
with the diffractively produced $\rho^0$ meson.

Let us finally turn to peripheral production of the $\pi^+\pi^0\pi^+\pi^-$
system. Of course, in this case it is of most interest to estimate the
contribution from the $\rho'^{\,+}$ intermediate state (see once again 
footnote 4). However, we succeeded in doing it only within the framework of 
the assumptions
which reduce to that the ratios $\sigma(\gamma p\to\rho'{\,^0}p)/\sigma(\gamma 
p\to\rho^0p)$ and $\sigma(\gamma p\to\rho'{\,^+}n)/\sigma(\gamma p\to\rho^+n)$
are taken to be approximately equal of each other. Let it be the case.
We shall also be guided by the following values: at $E_\gamma\approx6$ GeV, $
\sigma(\gamma p\to\rho^0p)\approx15\,\mu$b [74], $\sigma(\gamma N\to\rho^\pm N)
\approx0.58\,\mu$b (which is dominated by the $\rho$ exchange) [57,58], and $
\sigma(\gamma p\to\rho'{\,^0}p\to\pi^+\pi^-\pi^+\pi^-p)\approx1.5\,\mu$b (see
the beginning of this section). Hence, $\sigma(\gamma p\to\rho'{\,^0}p\to\pi^+
\pi^-\pi^+\pi^-p)/\sigma(\gamma p\to\rho^0p)\approx1/10$, and according to our
prescription \begin{equation}\sigma(\gamma p\to
\rho'{\,^+}n\to\pi^+\pi^0\pi^+\pi^-n)\approx\frac{1}{10}\,\sigma(\gamma p\to
\rho^+n)\times\left\{\begin{array}{c}1\\3/2\end{array}\right.\approx(0.058-0.08
7)\,\mu\mbox{b},\end{equation} where the factors 1 and 3/2 correspond to the
$\rho'\to\rho\sigma\to4\pi$ ($\sigma$ denotes the state of the $S$-wave $\pi\pi
$ system with $I=0$) and $\rho'\to a_1\pi\to4\pi$ decay models respectively.
Now we briefly discuss the input assumptions which lead to the relation $\sigma
(\gamma p\to\rho'^{\,0}p)/\sigma(\gamma p\to\rho^0p)\approx\sigma(\gamma p\to
\rho'^{\,+}n)/\sigma(\gamma p\to\rho^+n)$. There are two assumptions: 1)
diagonal vector dominance for the Regge exchange amplitudes with the vacuum and
non-vacuum quantum numbers in the $t$ channel in the reactions $\gamma N\to\rho
N$ and $\gamma N\to\rho'N$ and 2) universality of the $\rho$-Reggeon coupling
to hadrons. The diagonal vector dominance assumptions, as applied to the
reactions $\gamma p\to\rho^0p$ and $\gamma p\to\rho'^{\,0}p$, has been
discussed, for example, in Refs. [32,34,35,74,75]. In fact, there has been
shown in these works that such an approximation, with a glance to the quite
natural relations $\sigma_{tot}(\rho N)\approx\sigma_{tot}(\rho'N)\approx\sigma
_{tot}(\pi N)$, gives a reasonable explanation of the cross section value 
observed for $\gamma p\to\rho'^{\,0}p$. On the other hand, the absence of some
evidence
for the $\rho'\to\rho\rho$ decay permits the diagonal vector dominance model to
be also applied to the vertex $\gamma(\rho^+)\rho'^{\,+}$, where $(\rho^+)$ is
a Reggeon. Adding to this the assumption 2) about $\rho$-universality, i.e.,
about the approximate equality of the $\rho^0(\rho^+)\rho^+$ and $\rho'^{\,0}(
\rho^+)\rho'^{\,+}$ vertices, we come to Eq. (18).

It should be noted that the expected suppression of $\sigma(\gamma
p\to\rho'^{\,+}n\to\pi^+\pi^0\pi^+\pi^-n)$ relative to $\sigma(\gamma p\to
\rho'^{\,0}p\to\pi^+\pi^-\pi^+\pi^-p)$ is a major reason why we suggest to look
for in photoproduction the $X^\pm$ states rather than the $X^0$ one \footnote{
In this connection, we also point out the relation $\sigma(\gamma N\to X^0N\to
\rho^0\rho^0N)/\sigma(\gamma N\to X^\pm N\to\rho^\pm\rho^0N)=4/9$ and the
possibility of the additional rich background in the $\rho^0\rho^0\to\pi^+\pi^-
\pi^+\pi^-$ channel from the $I=0$ states.}.

Comparing Eqs. (12) and (18), we conclude that if one succeeds in selecting the
events due to peripheral production of the $\pi^\pm\pi^0\pi^+\pi^-$ system in
$\gamma N\to\pi^\pm\pi^0\pi^+\pi^-N$, then the separation of the $\rho'^{\,\pm}
$ and $X^\pm$ contributions would be quite possible with high statistics. In so
doing, a detailed simultaneous analysis of all two- and three-pion mass spectra
(for example, $\pi^+\pi^0$, $\pi^+\pi^-$, $\pi^+\pi^+$, $\pi^0\pi^-$, $\pi^+\pi
^+\pi^0$, $\pi^+\pi^+\pi^-$ and $\pi^+\pi^-\pi^0$ in $\gamma p\to\pi^+\pi^0\pi^
+\pi^-n$) and corresponding angular distributions has to come into play. Note
that the simulation of the mass and angular distributions for the decays $X\to
\rho\rho\to4\pi$ has been described in detail in the literature [12-17].
   A detectable reduction of the background from $\rho
   '^{\,\pm}$ production and, consequently, a much more accurate separation of
   the $\rho'^{\,\pm}$ and $X^\pm$ signals can be provided by using polarized
   photons. As is known, in this case, good analyzers for $4\pi$ states are
   the vector $\vec Q=\vec p_{\pi_1}+\vec p_{\pi_2}$ (where $\pi_1$ and $\pi_2
   $ are equally charged pions) and the photon polarization angle $\psi$ (for
   reviews see e.g. [31,73,76,77]). One can make sure that within the simplest
   production and decay models the $(\vec Q,\, \psi)$-distributions for the
   $\rho'^{\,\pm}$ and $X^\pm$ states are essentially different. Certainly,
   because of the great extent, it is more appropriate to carry out the full
   analysis of the use from polarized photons elsewhere.
   
As for the $\rho_3(1690)$ resonance, no analysis has so far been done including
its contribution to the description of the observed peaks seen in 
photoproduction
of $\pi^+\pi^-\pi^+\pi^-$, $\pi^+\pi^0\pi^-\pi^0$ and $\pi^+\pi^-$ which have
been attributed to the $\rho'$ resonance [28]. However, diffractive production
of $\rho^0_3$ in the $a_2^\pm\pi^\mp$ decay modes was found in the reaction
$\gamma p\to\eta\pi^+\pi^-p$ for $20<E_\gamma<70$ GeV [28] and, on the basis of
these data, the cross sections $\sigma(\gamma p\to\rho^0_3p\to\pi^+\pi^-\pi^+
\pi^-p)=0.147\pm0.42\pm0.32\ \mu$b and $\sigma(\gamma p\to\rho^0_3p\to\rho^+
\rho^-p)=0.018\pm0.016\pm0.004\ \mu$b were predicted as well. Furthermore,
before the $\rho'$ state had been discovered, some upper limits for the $\rho
^0_3$ production cross sections were determined, namely, $\sigma(\gamma p\to
\rho^0_3p
\to\pi^+\pi^-p)\leq0.85\pm0.35\ \mu$b for $E_\gamma=4.3$ GeV [54], $\sigma(
\gamma p\to\rho^0_3p\to\pi^+\pi^-p)<0.1\ \mu$b and $\sigma(\gamma p\to\rho^0_3p
\to\pi^+\pi^-\pi^+\pi^-p)<1\ \mu$b for $E_\gamma=5.25$ GeV [53]. The results of
Refs. [53,54] have been based on very poor statistics and at present it is not 
clear 
whether they have much to do with the $\rho^0_3(1690)$. In this situation, a
reliable estimate of the cross section for the charge exchange reaction $
\gamma p\to\rho^+_3n\to\pi^+\pi^0\pi^+\pi^-n$ is rather difficult to obtain.
If there occurs an universal relation between the Pomeron contribution and the
$f_2$ Regge pole contribution in the reactions $\gamma p\to\rho^0p$ and $
\gamma p\to\rho
^0_3p$ and if exchange degeneracy [46] and the naive quark counting rules hold
for the $\rho$, $a_2$, and $f_2$ exchanges in $\gamma N\to\rho_3N$, then it is
hoped that $\sigma(\gamma p\to\rho^+_3n)$ would be an odder of magnitude
smaller than $\sigma(\gamma p\to\rho'^{\,+}n)$.

Judging by the data presented in Table 2, $\sigma(\gamma N\to\pi^\pm\pi^0\pi^+
\pi^-N)\approx7\ \mu$b at $E_\gamma\approx6$ GeV. We have shown that a part of
the cross section of about 3 $\mu$b can be explained by a few processes with
the appreciable cross sections due to rather simple mechanisms (see Eqs. $(15)-
(18)$). The remainder of the total cross section is probably distributed among
a great many of more ``fine" channels a set of which has been indicated by us
only partly. In this sense, the presented analysis is preliminary and further
theoretical investigations are needed. Undoubtedly, a further crucial progress
in refinement of the cross sections for various photoproduction channels
will be connected with the accurate measurements at JLAB. It is
for experimentalists to decide.

Let us now consider very briefly the reaction $\gamma p\to\pi^-\pi^0\pi^+\pi^-
\Delta^{++}$. The available data for $E_\gamma<10$ GeV are presented in Table
III. According this information, the most probable value of $\sigma(\gamma p\to
\pi^-\pi^0\pi^+\pi^-\Delta^{++})$ for $4\leq E_\gamma\leq6$ GeV is about $1.87
\pm0.38\ \mu$b. As 
is seen from Table III, the only observable partial channel $\gamma p\to\omega
\pi^-\Delta^{++}$ [53,54] can contribute to the $\pi^-\pi^0\pi^+\pi^-\Delta^{++
}$ production cross section up to 1 $\mu$b. As for the estimate of the cross
section for the channel $\gamma p\to\rho'{\,^-}\Delta^{++}\to\pi^-\pi^0\pi^+
\pi^-\Delta^{++}$, it can be obtained by multiplying the cross section value
in Eq. (18) by the coefficient 1.75. Similarly we already have done by passing
from Eq. (12) to Eq. (14) in the case of $X^-$ production. We thus expect $
\sigma(\gamma
p\to\rho'{\,^-}\Delta^{++}\to\pi^-\pi^0\pi^+\pi^-\Delta^{++})\approx(0.1-0.15)
\ \mu$b. In principle, the reaction involving $\Delta^{++}$ production may be
found to be rather favorable (in the sense of the background conditions) to the
search for the $X^-$ signal.

In conclusion we note that the reaction $\gamma p\to\omega\pi^-\pi^+p$ may be
dominated by associated $\omega\pi^-\Delta^{++}$ production (see Table III). 
This point has been especially emphasized, for example, in Ref. [54]. The 
cross section for $\gamma 
p\to\omega\pi^-\pi^+p$ with the $\omega\pi^-\Delta^{++}$ channel excluded 
is not likely to be over $(0.5-1)\ \mu$b at $E_\gamma\approx6$ GeV (see 
Table III). In our opinion, peripheral production of the $\omega\pi^-\pi^+$
systems is of special interest in the reaction $\gamma p\to\omega\pi^-\pi^+p$.
In particular, the $\omega(1600)$ and $\omega_3(1670)$ resonances have to be
photoproduced diffractively in the $C$-odd $\omega\pi^-\pi^+$ states with the
cross sections an order of magnitude smaller then are now available (or
expected) for $\gamma p\to\rho'^{\,0}p$ and $\gamma p\to\rho^0_3p$ respectively
(i.e., for example, $\sigma(\gamma p\to\omega(1600)p\to\omega\pi^-\pi^+p)
\approx0.1\ \mu$b). To search for the $C$-even resonances, the $\omega\rho^0$
mode is most suitable. In fact, of the known (``tabular") $q\bar q$ resonances 
in this decay mode there are only the $a_2(1320)$ [26,78] and, possibly,
$\pi_2(1670)$ and $\pi(1800)$ states [78]. 
Therefore, the appreciable enhancement with $(J^P,|J_z|)=(2^+,2)$ 
found by the ARGUS group in the cross section of the reaction
$\gamma\gamma\to\omega\rho^0$ for $1.5\leq W_{\gamma\gamma}\leq2.1$ GeV [79]
may be interpreted as evidence for a $q^2\bar q^2$ MIT-bag state
which is strongly coupled to the $\omega\rho^0$ channel [20,17]. In the
notations of Ref. [20], it is the $C^0_\pi(36)$ state with $I^G(J^{PC})=1^-(2^
{++})$ being the partner of the $X(1600,2^+(2^{++}))$. Note, without 
going into detailed estimates, that we would expect the cross section for
$\gamma p\to C^0_\pi(36)p\to\omega\rho^0p$ at a level of about 0.075 $\mu$b at
$E_\gamma\approx6$ GeV.
This value should be compared with the estimate given by Eq. (12). The expected
enhancement of the $C^0_\pi(36)$ photoproduction cross section by comparison
with the case of $X^\pm$ is due to several factors. For example, one of them is
a strong coupling of the $\omega$ exchange (which we consider as a major
mechanism of the reaction $\gamma p\to C^0_\pi(36)p$) to the nucleons. Also, 
we would expect $\sigma(\gamma p\to C^+_\pi(36)n\to\omega\rho^+n)\approx(5
/81)\sigma(\gamma p\to C^0_\pi(36)p\to\omega\rho^0p)\approx0.0046$ $\mu$b and,
for $C^s_\pi(36)$ and $C^s(36)$ MIT-bag state production, $\sigma(\gamma p\to
C^{s+}_\pi(36)n\to\phi\rho^+n)\approx2\sigma(\gamma p\to C^{s0}_\pi(36)p\to\phi
\rho^0p)\approx(10/81)\sigma(\gamma p\to C^0_\pi(36)p\to\omega\rho^0p)B(C^s_\pi
(36)\to\phi\rho)\approx(0.0092\,\mu$b$)B(C^s_\pi(36)\to\phi\rho)$ and $\sigma(
\gamma p\to C^s(36)p\to\phi\omega p)\approx(2/9)\sigma(\gamma p\to C^0_\pi(36)p
\to\omega\rho^0p)B(C^s(36)\to\phi\omega)\approx(0.0166\,\mu$b$)B(C^s(36)\to\phi
\omega)$. A search for the threshold enhancements in the $K^*\bar K^*$ and $
\omega\omega$ final states in photoproduction would also be an important
complement to the $\gamma\gamma$ experiments and to other approaches. Note 
that it is not yet clearly established that some threshold enhancements in $
\gamma\gamma\to VV'$ are due to exotic meson resonances.
 \vspace*{1cm}
\begin{center} Table I. \ Classification of the $\rho^\pm\rho^0$ states.
\end{center} \begin{center}\begin{tabular}{|c|c|c|} \hline
$\,I^G(J^{PC})\,$ series,\, & Possible resonance & $^{2S+1}L_J$ configurations
for \\ $k=0,1,2,...$ & states   & lower $J$ states  \\ \hline
$1^+((2k+1)^{--})$ & $\rho(1700)$, \ $\rho_3(1690)$ & ($^1P_1$, \,$^5P_1$, \,$
^5F_1$); \,($^5P_3$, \,$^1F_3$, \,$^5F_3$, \,$^5H_3$) \\
$1^+((2k+1)^{+-})$ & $b_1(1235)$                & $^3S_1$, \,$^3D_1$ \\
$1^+((2k+2)^{--})$ & $\rho_2$(?)                & $^5P_2$, \,$^5F_2$ \\
$1^+((2k+2)^{+-})$ & are absent in $q\bar q$ system & $^3D_2$        \\ 
$2^+((2k+1)^{-+})$ & $q^2\bar q^2$              & $^3P_1$, \,$^3F_1$ \\
$2^+((2k+1)^{++})$ & $q^2\bar q^2$              & $^5D_1$         \\
$2^+((2k)^{-+})$ &  $q^2\bar q^2$               & ($^3P_0$); \,($^3P_2$) \\
$2^+((2k)^{++})$ &  $\ q^2\bar q^2$, $X(1600, 2^+(2^{++}))\ $
& ($^1S_0$, \,$^5D_0$); \,($^5S_2$, \,$^1D_2$, \,$^5D_2$, \,$^5G_2)\ $\\
\hline \end{tabular} \end{center} \vspace*{0.5cm}
\begin{center} Table II. Total and partial cross sections of the reactions
$\gamma N\to\pi^\pm\pi^0\pi^+\pi^-N$.
\end{center} \begin{center} \begin{tabular}{|c|c|c|c|c|c|} \hline
$E_\gamma$ & Reaction &  Cross  & $E_\gamma$ & Reaction (Refs. &  Cross  \\
(GeV) & (Ref. [62]) & section & (GeV) & [61,63,68,70-72]) & section\\
      &   & ($\mu$b) &   &                       & ($\mu$b)\\ \hline
4.3       & $ \gamma n\to\pi^-\pi^0\pi^+\pi^-p $ & $7.5\pm1.0$   &
$6.9-8.1$ & $ \gamma n\to\pi^-\pi^0\pi^+\pi^-p $ & $4.85\pm0.89$    \\
          & $ \gamma n\to\omega\pi^-p          $ & $1.4\pm0.5$   &
$3.6-5.1$ & $ \gamma n\to\pi^-\pi^0\pi^+\pi^-p $ & $11.0\pm2.2$     \\
          & $ \gamma n\to\rho^-\pi^+\pi^-p     $ & $1.1\pm0.5$   &
7.5       & $ \gamma n\to\pi^-\pi^0\pi^+\pi^-p $ & $6.1\pm0.8$      \\
          & $ \gamma n\to\rho^0\pi^0\pi^-p     $ & $1.8\pm1.0$   &
$2.5-5.3$ & $ \gamma n\to\omega\pi^-p          $ & $1.6\pm0.5$      \\
          & $ \gamma n\to\rho^+\pi^-\pi^-p     $ & $0.5\pm0.5$   &
$4.2-4.8$ & $ \gamma p\to\omega\Delta^+\to\omega\pi^+n $ & $0.83\pm0.10$ \\
          & $ \gamma n\to\pi^+\pi^-\pi^0\Delta^0 $ & $0.6\pm0.6$ &
8.9       & $ \gamma N\to\omega\Delta\to\omega\pi^\pm N $ & $0.24\pm0.023$ \\
\hline\end{tabular}\end{center} \vspace*{0.5cm}
\begin{center} Table III. Cross sections for the reactions $\gamma p\to\pi^
-\pi^0\pi^+\pi^-\Delta^{++}$ and $\gamma p\to\omega\pi^-\pi^+p$,\\
and for their common partial channel $\gamma p\to\omega\pi^-\Delta^{++}$.
\end{center} \begin{center} \begin{tabular}{|c|c|c|c|c|} \hline
$E_\gamma$ & $\sigma(\gamma p\to\pi^-\pi^0\pi^+\pi^-\Delta^{++})$ &
$\sigma(\gamma p\to\omega\pi^-\Delta^{++})$ & $\sigma(\gamma p\to\omega\pi^-
\pi^+p)$  & Refs. \\ (GeV) & ($\mu$b) & ($\mu$b) & ($\mu$b) &  \\ \hline
4.3       & $2.4\pm0.8 $      & $\approx1$  & $1.42\pm0.45$      & [54] \\
$4-6$     & $1.3\pm0.3\pm0.2$ & ---         & $1.6\pm0.2\pm0.24$ & [69] \\
$4.5-5.8$ & $\leq2.4\pm1.1$   & ---         & $2.4\pm0.9$        & [59] \\
5.25      & $3.9\pm1.5$       & $0.5\pm0.2$ & $1.5\pm0.4$        & [53] \\
\hline\end{tabular}\end{center} \newpage\vspace*{0.5cm} 
\begin{center} \large{\bf Figure caption}\end{center} 
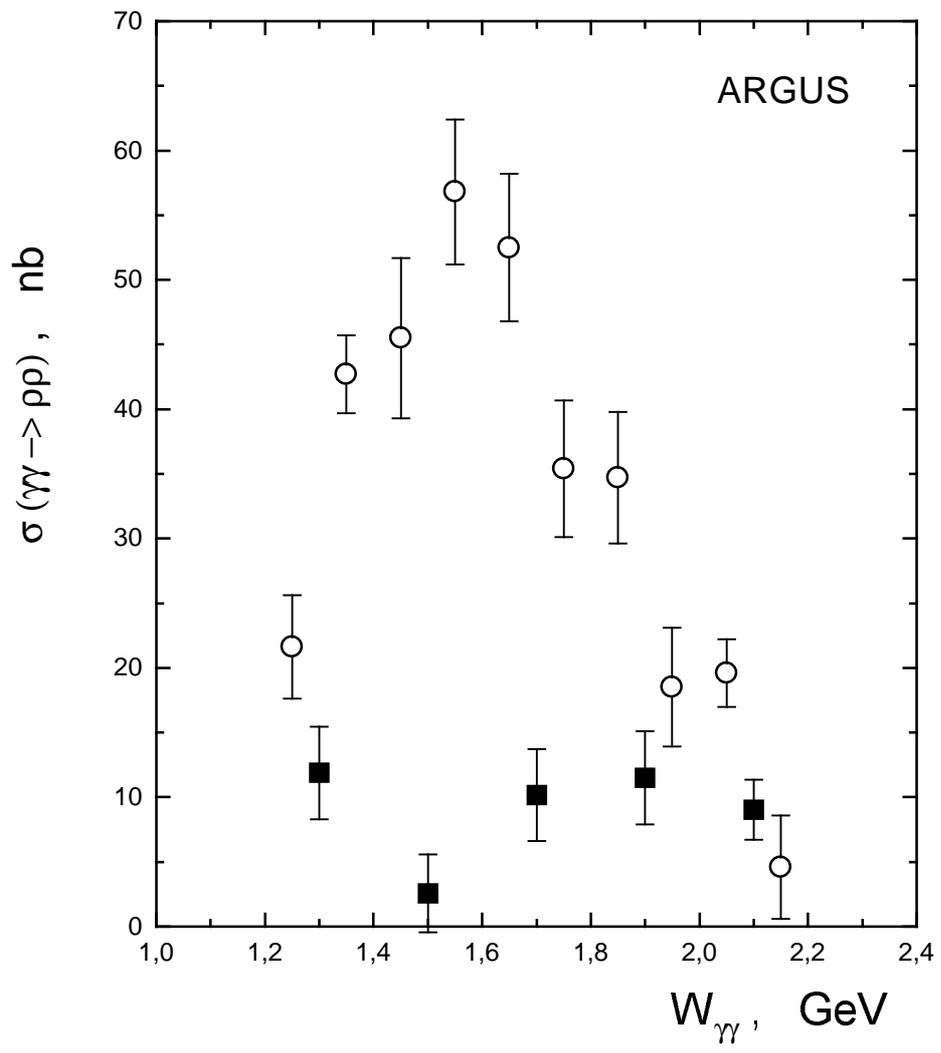
{\bf Fig. 1.} The ARGUS results on the $(\,J^P,\,J_z)=(2^+,\,\pm2)$ partial
cross sections for the reactions $\gamma\gamma\to\rho^0\rho^0$ [13] (open
circles) and $\gamma\gamma\to\rho^+\rho^-$ [15] (full squares).
$W_{\gamma\gamma}$ is the invariant mass of the $\gamma\gamma$ system. For an
ordinary isospin 0 resonance one expects $\sigma(\gamma\gamma\to\rho^+\rho^-)/
\sigma(\gamma\gamma\to\rho^0\rho^0)=2$ (and 1/2 for a pure isospin 2
resonance). Instead, the observed ratio is lower than 1/2. A resonance
interpretation for such a result in terms of $q^2\bar q^2$ states thus
requires the presence of a flavor exotic $I=2$ resonance which interferes with
isoscalar contributions [11,17,18].
\end{document}